# Ground State Study of a Double Core-Shell Dendrimer Nanostructure

H. Labrim[1], A. Belhaj[2], S. Ziti[3], L. Bahmad[4*], A. Elmorabiti[1], L. Laânab[5], S. Benmokhtar[6], A. Benyoussef[4]

[1] Centre National de l'Energie, des Sciences et des Techniques Nucléaires, Rabat, Morocco.

[2] Département de Physique, Faculté Poly-disciplinaire, Université Sultan Moulay Slimane, Béni Mellal, Morocco

[3] Laboratoire de Magnétisme et Physique des Hautes Énergies (LMPHE-URAC 12), Université Mohammed V, Faculté des Sciences, B.P. 1014, Rabat, Morocco.

[4] Labortoire de Recherche en Informatique, Université Mohammed V, Rabat, Morocco.

[5] Laboratoire Conception et Systèmes (LCS), Université Mohammed V, Faculté des Sciences, B.P. 1014, Rabat, Morocco.

[6] Laboratoire de Chimie Physique des Matériaux LCPM, Faculté des Sciences Ben M'Sik, Casablanca, Morocco.

**Abstract**

Based on a dendrimer graph theory, we investigate the ground state phase diagrams of a nanostructure formed by three spin types: $\sigma=1/2$, $S=1$ and $q=3/2$. This nanostructure is built on a dendrimer with triangular core geometry. More precisely, we evaluate the effect of different physical parameters including the coupling exchange interactions, the external magnetic and the crystal fields. By varying these parameters, we analytically elaborate the ground state phase diagrams. Concretely, we found that the existence of the crystal field increases the number of stable phases. The total number of possible phases is 24=2x3x4, which are : $(\pm 1/2, \pm 1, \pm 3/2)$, $(\pm 1/2, 0, \pm 3/2)$, $(\pm 1/2, \pm 1, \pm 1/2)$ and $(\pm 1/2, 0, \pm 1/2)$. All these phases are found to be stable expect two special phases: $(+1/2, 0, -3/2)$ and $(-1/2, 0, +3/2)$, which are masked by the other phases in all the ground state phase diagrams.

# 1. Introduction

Several studies of magnetic physical properties of strongly correlated electron models have been carried out, in connection with the elaboration of nano-materials using different calculation methods with appropriate approximations [1-9]. It is recalled that the hexagonal structure is the most stable one which has been explored to elaborate new materials involving new hexagonal symmetry. In fact, many models have been investigated using different methods and approaches.

Recently, a special emphasis put on the study of the interaction of metal atoms placed on various structures [10]. These interactions can be used for many potential applications like nano-magntic materials and spintronic. Several works have been focused on studying such models using first-principles density functional theory (DFT) [11,12]. The corresponding studied is usually used in experiments and many other related topics. In fact, many physical quantities have been computed including the Curie temperature and the magnetic moments per formula, the energy adsorption, geometry and the density of states [13-15]. Double geometrical structure has been also investigated connection with symmetries in physics. A close inspection reveals that the $G_2$ hexagons appearing in the classification of Lie algebras have explored to build new materials like grapheme [16,17]. In this way, the principal hexagonal unit cell contains now twelve atoms leading to $(1\times1)$ and $(\sqrt{3}\times\sqrt{3})R30^0$ superstructures on the two parallel sheets separated by a length distance $\Delta$ fixed by the relaxation method. Using WIEN2K code based on DFT method, the electronic and the magnetic properties of double hexagonal material have been worked out [18-19].

More recently, the dendrimer graph theory has been used in the study of spin models [20,21]. The aim of this paper is contribute to this activity by exploring dendrimer graph theory to investigate the ground state phase diagrams of a nanostructure formed by three spin types which are $\sigma$=1/2, S=1 and q=3/2. This nanostructure is built from a dendrimer with triangular core geometry. Concretely, we evaluate the effect of different physical parameters including the coupling exchange interactions, the external magnetic and the crystal fields. Varying such parameters, we analytically elaborate the ground state phase diagrams. More precisely, we found that the existence of crystal field increases the number of stable phases which are (1/2,1,3/2), (-1/2,-1,-3/2), (-1/2,1,3/2), (1/2,-1,-3/2), (-1/2,1,-3/2), (1/2, -1, 3/2), (-1/2,-1,3/2) and (1/2,1,-3/2).

## 2. Double Core shell Dendrimer model

In this part, we build a nano-structure inspired by dendrimer graph theory. To do so, we first construct the studied structure which will be based on double triangular core-shell geometry. It is worth noting that the dendrimer structures have been explored in many biological works including DNA activities. To manipulate the corresponding structure, one may use techniques based on graph theory using vertices and links forming nice quivers. A close inspection reveals that there are many graphical configurations. For simplicity, we consider a special geometry based on a Cayley tree graph theory. The geometry will be used here is illustrated in Fig 1.

The system, that we study, is formed by three different spin particles: $\sigma = 1/2$, S=1 and q=3/2 arranged according to Cayley tree geometry, given in Fig. 1. In particular, the q spins are placed in the external surface whereas the spins $\sigma$ and S are located in the core and intermediate surfaces, respectively. The corresponding Hamiltonian reads as

$$\mathcal{H} = -J_1 \sum_{<i,j>} \sigma_i \sigma_j - J_2 \sum_{<k,l>} S_k \sigma_l - J_3 \sum_{<m,n>} S_m q_n - \Delta \sum_i (S_i^2 + q_i^2) - H \sum_i (S_i + \sigma_i + q_i) \quad (1)$$

Here, <i,j>, <k,l>, <m,n> stands for nearest-neighbors between $\sigma - \sigma$, $S - \sigma$ and $S - q$ spins, respectively. The parameters $J_1$, $J_2$ and $J_3$ are the exchange coupling constants between $\sigma - \sigma$, $S - \sigma$ and $S - q$ interactions, respectively. H stands for an external magnetic field, and $\Delta$ represents a crystal field applied on the all system spins. In all the following, we will fix the exchange coupling constant $J_1$ to its unit value.

## 3. Ground state phase diagrams

The ground state phase diagrams are explored by computing and comparing all possible configuration energies of the spins $N_\sigma \times N_S \times N_q = 2 \times 3 \times 4 = 24$. Where, $N_\sigma$ denotes the number of spins of $= \pm \frac{1}{2}$, $N_S$ is the number of states of S=0, $\pm 1$ and $N_q$ corresponds to the number of states of q= $\pm \frac{1}{2}, \pm \frac{3}{2}$.

The aim of this work is determine the corresponding ground phase diagrams by varying different physical parameters controlling the Hamiltonian (1). Analytically, we elaborate the ground state phase diagrams for the more stable configurations by minimizing the energies produced by the Hamiltonian (1) in terms of adequate parameters. The corresponding phase diagrams are presented in Figs. 2(a)-(f).

Fig. 2a presents the most stable phases in the plane ($J_3$, $J_2$), in the absence of any crystal and external magnetic fields H=0 and $\Delta$ =0 for fixed constant coupling $J_1$=1. From this figure, it is found that the only eight (8) phases are stable, which are: ($\pm 1/2$, $\pm 1$, $\pm 3/2$). Such phases correspond to the maximal amplitudes of σ, S and q. The phases σ and S are not sensible to increasing values of the exchange coupling $J_3$. However, the q stable phases depend strongly on the sign of this parameter. On the other hand, the phases of σ are not affected when varying the exchange coupling $J_2$. However, the S and q stable phases are strongly affected when changing the sign of such parameter.

In Fig. 2b, we present the most stable phases in the plane (H, $J_2$), in the absence of any crystal field $\Delta$ =0 for fixed values of the exchange constant couplings $J_1$= $J_3$=-1. From this figure, it is found that the most stable phases corresponding the maximal amplitudes of σ, S and q are also found to be stable. The external magnetic field separates symmetrically these phases. The separation is due the $Z_2$ symmetry acting on such phases via the following relation scheme

$$(x, y, z) \rightarrow (-x, -y, -z)$$

The increasing effect of the exchange constant coupling $J_2$ does not affect the states of S and q, but affects strongly the states of σ, see Fig. 2b.

Fig. 2c, illustrates the most stable phases in the plane (H, $J_3$), in the absence of any crystal field $\Delta$ =0 and fixed constant coupling values $J_2$= $-J_1$=-1. From this figure, it follows that there is $Z_2$ symmetry, associated with the external magnetic field h. Only six (6) stable phases, from eight, are found to be stable when compared with Fig. 2(a,b). These phases are: (1/2, 1, 3/2), (-1/2,-1,-3/2), (-1/2, 1, 3/2), (1/2, -1, -3/2), (-1/2,1,-3/2) and (1/2, -1, 3/2). The phases (-1/2,-1,3/2) and (1/2,1,-3/2), do not appear due to effect of negative fixed value of the exchange coupling parameter $J_2$=-1.

In order to investigate the crystal field effect on the stable phases, in the absence of any magnetic field, H=0, we present in the planes ($\Delta$, $J_2$) and ($\Delta$, $J_3$) the corresponding ground state phase diagrams. This is illustrated by Figs. 2(d) for $J_3$=-1.0; and 2(e) for $J_2$=-1.0, respectively. It is found that each phase diagram shows 12 stable phases. But only eight stable phases are common between these two phase diagrams, namely : (1/2,-1,-3/2), (-1/2,1,-3/2), (1/2,0,1/2), (1/2,0,-1/2), (-1/2,0,1/2), (-1/2, 0, -1/2), (1/2,-1,1/2) and (-1/2,1,-1/2).

In order to inspect the effect of both the crystal and the magnetic field effect, we illustrate in Fig. (2f) the associated phase diagram in the plane (H, $\Delta$) for $J_2$=-1.0 and $J_3$=-1.0. From this figure, it is found that four new phases are stable for negative values of the crystal field $\Delta$. These phases are (1/2,1,1/2), (-1/2,-1,-1/2), (1/2,0,3/2) and (-1/2,0,-3/2). The presence of the phases containing S=0 and q=±1/2 is due to the fact that these states are minimizing the energy given by the Hamiltonian (Eq. 1). For positive values of the crystal field $\Delta$, the two phases (1/2,1,3/2) and (-1/2,-1,-3/2), are not affect by the increasing the crystal field effect. Table 1 summarizes the all stable phases. They are 22 stable phases, from 24 possible phases, which can be found in different ground state phase diagrams.

|  | Fig. 2(a) | Fig. 2(b) | Fig. 2(c) | Fig. 2(d) | Fig. 2(e) | Fig. 2(f) |
|---|---|---|---|---|---|---|
| (+1/2,+1,+3/2) | X | X | X |  |  | X |
| (-1/2,-1,-3/2) | X | X | X |  |  | X |
| (1/2,-1,-3/2) | X | X | X |  | X |  |
| (-1/2,1,3/2) | X | X | X |  | X |  |
| (1/2,1,-3/2) | X | X |  | X |  |  |
| (-1/2,-1,3/2) | X | X |  | X |  |  |
| (1/2,-1,3/2) | X | X | X | X | X | X |
| (-1/2,1,-3/2) | X | X | X | X | X | X |
| (1/2,0,1/2) |  |  |  | X | X | X |
| (1/2,0,-1/2) |  |  |  | X | X |  |
| (-1/2,0,1/2) |  |  |  | X | X |  |
| (-1/2,0,-1/2) |  |  |  | X | X | X |
| (1/2,-1,1/2) |  |  |  | X | X | X |
| (-1/2,1,-1/2) |  |  |  | X | X | X |
| (1/2,1,-1/2) |  |  |  | X |  |  |
| (-1/2,-1,1/2) |  |  |  | X |  |  |
| (1/2,-1,-1/2) |  |  |  |  | X |  |
| (-1/2,1,1/2) |  |  |  |  | X |  |
| (1/2,0,3/2) | | | | | | X |
| (-1/2,0,-3/2) | | | | | | X |
| (1/2,1,1/2) | | | | | | X |
| (-1/2,-1,-1/2) | | | | | | X |

Table 1: Evolution of the 22 (from 24) possible stable phases in the different ground state phase diagrams, shown in Figs. 2(a)-(f). The two special phases (+1/2,0,-3/2) and (-1/2,0,+3/2) are the only absent phases in these phase diagrams. Fig. 2(a) is plotted in the plane ($J_3$, $J_2$), Fig. 2(b) corresponds to the plane (H, $J_2$), Fig. 2(c) in in the plane (H, $J_3$), Fig. 2(d) in in the plane ($\Delta$, $J_2$), Fig. 2(e) in in the plane ($\Delta$, $J_3$), whereas Fig. 2(f) is plotted in the plane (H, $\Delta$).

## 4. Conclusions

In this work, we have used dendrimer graph theory to study of ground state phase diagrams of a nanostructure based on a double core sell dendrimer geometry. The model has been formed with : **σ** =1/2, S=1 and q=3/2 spins arranged according to Cayley tree geometry, see Fig. 1. In particular, the q spins are placed in the external surface whereas the spins **σ** and S are located in the core and intermediate surfaces, respectively. We have examined the effect of the coupling exchange interactions in the absence/presence of both an external magnetic and crystal fields. More precisely, we have elaborated the ground state phase diagrams. Then, we have discussed the stable phases. In particular, we have found that the presence of the crystal fiend increases the number of stable phases which are (1/2,1,3/2), (-1/2,-1,-3/2), (-1/2,1,3/2), (1/2,-1,-3/2), (-1/2,1,-3/2), (1/2, -1, 3/2), (-1/2,-1,3/2) and (1/2,1,-3/2).

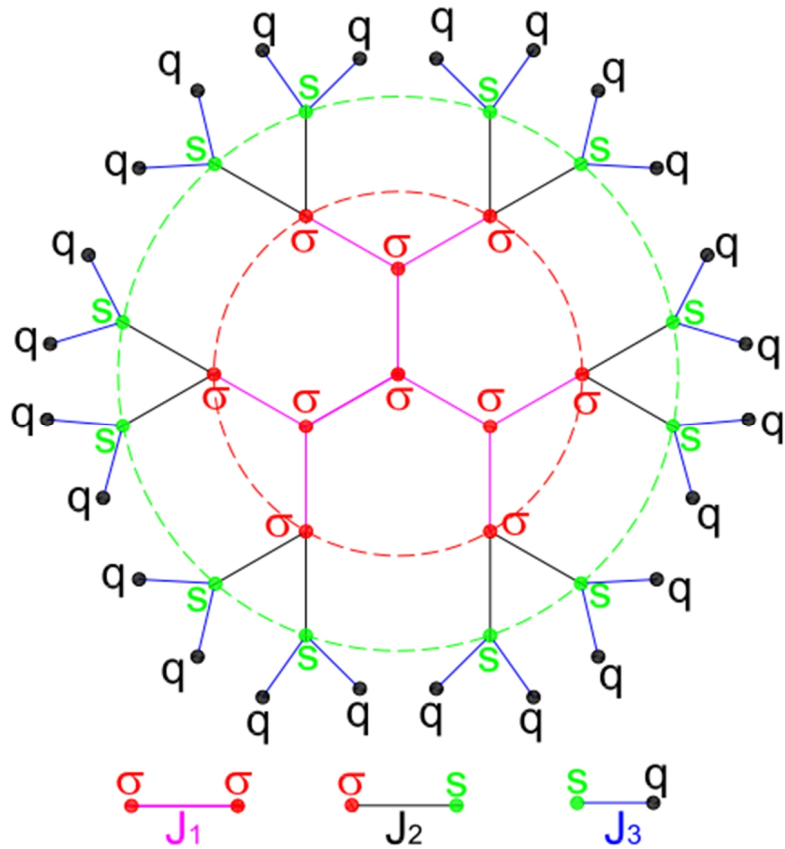

Fig. 1: A sketch of the geometry of the nanostructure based on a dendrimer spins system. It is formed by a fixed number of mixed **σ** and **S**–spins : $N_{Tot}=N_\sigma+N_S+N_q$ spins, with $N_\sigma=10$, $N_S=12$ and $N_q=24$ spins.

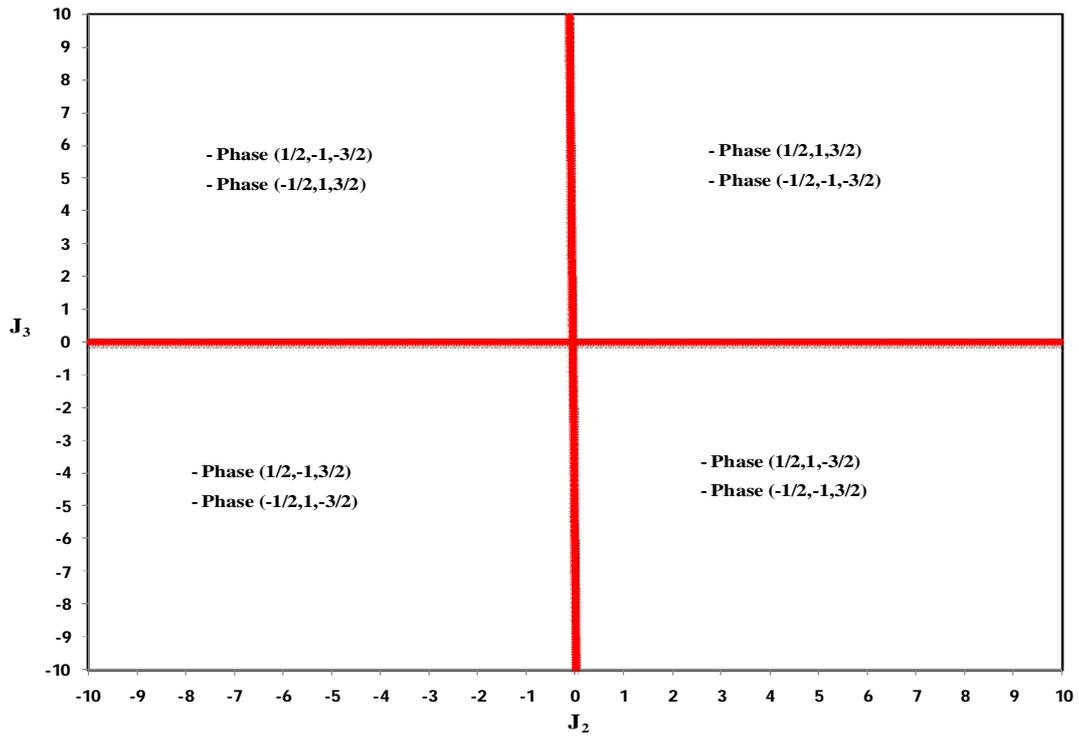

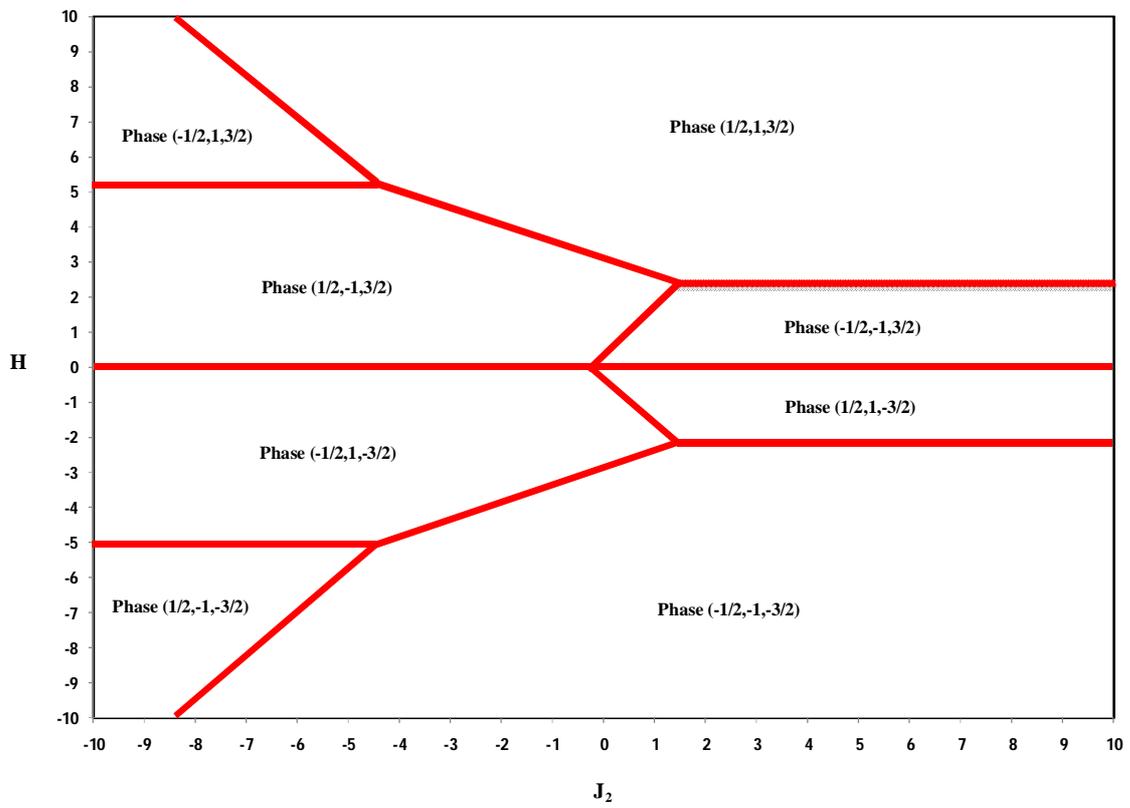

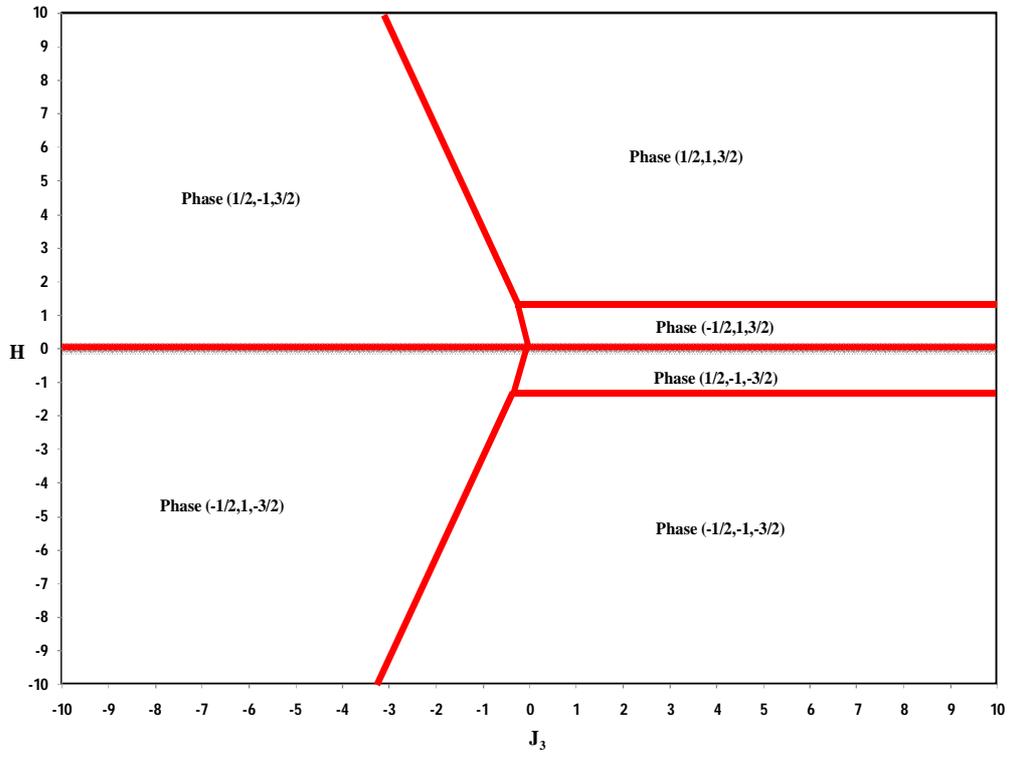

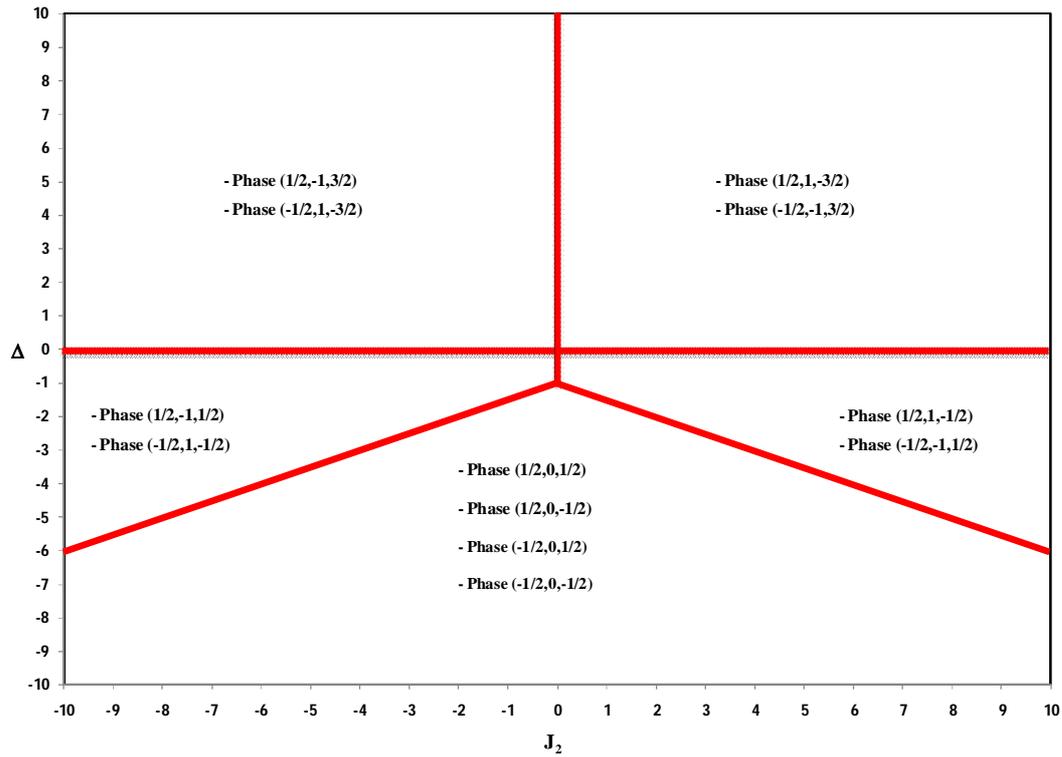

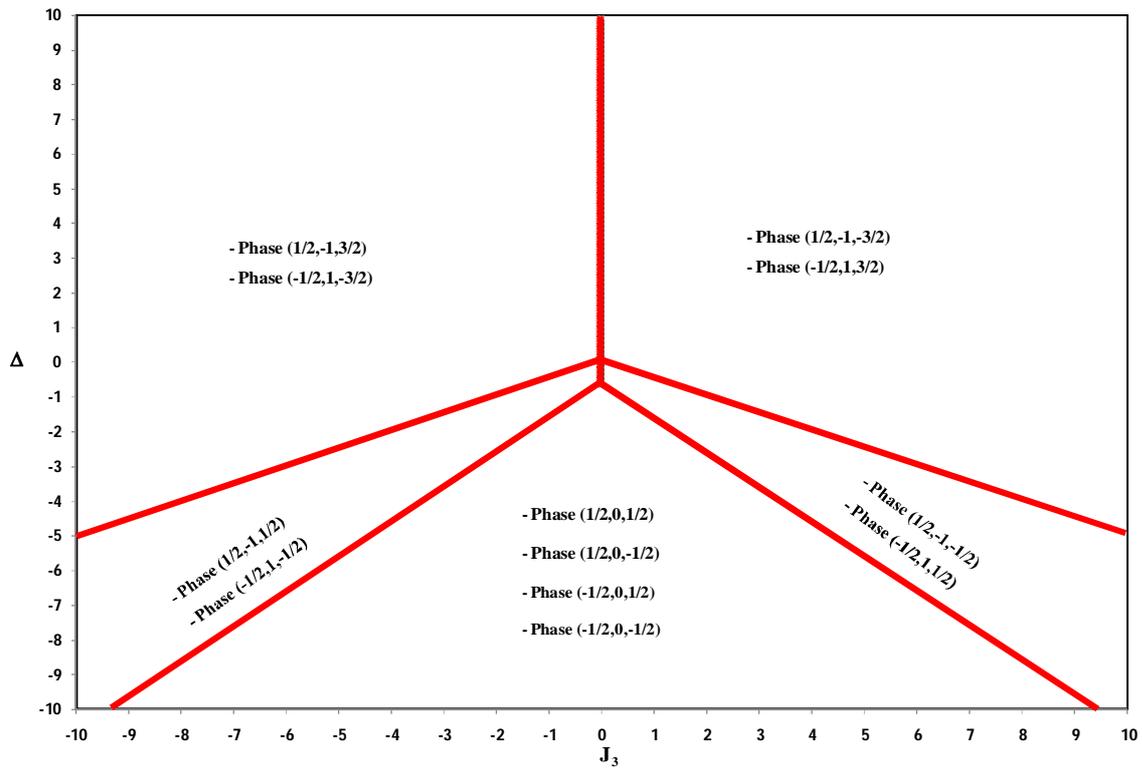

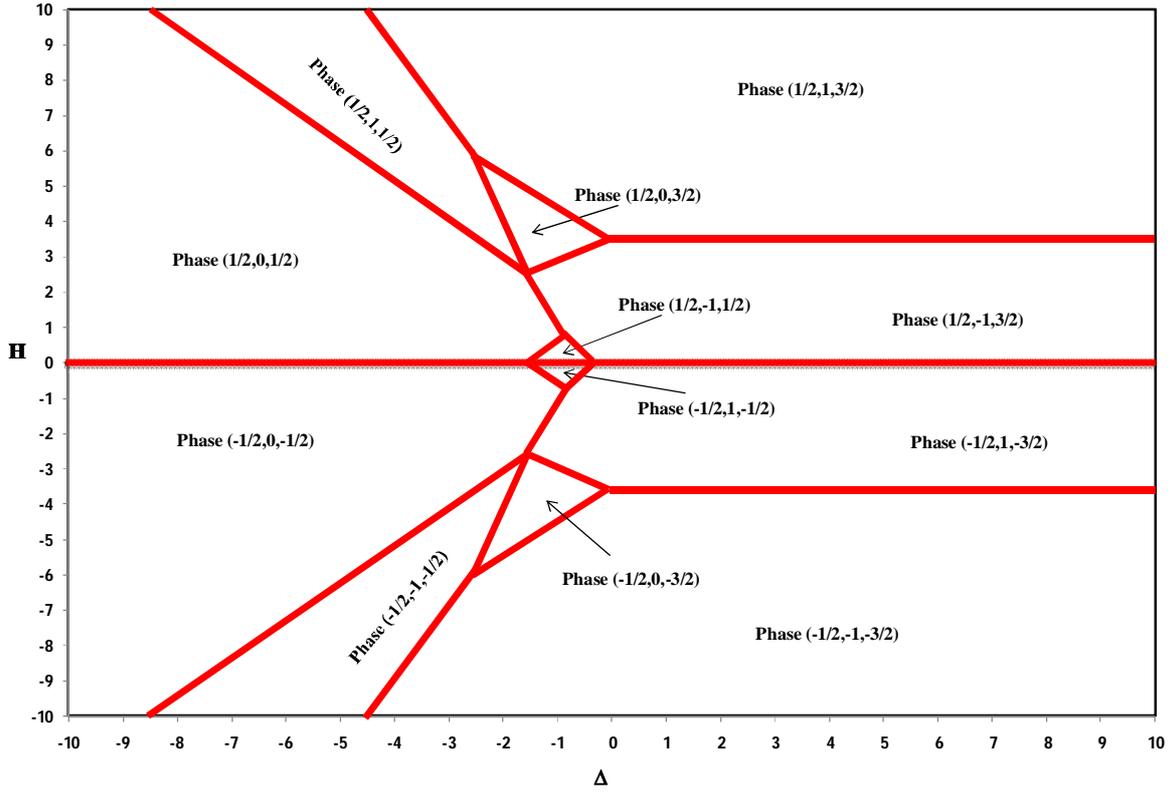

**Fig. 2:** Ground state phase diagrams, showing different stable phase in different planes : (a) in the plane ($J_1$, $J_2$) for H=0.0, Δ=0.0; (b) in the plane (H, $J_2$) for Δ=0.0, $J_1$=+1.0, $J_3$=-1.0; (c) in the plane (H, $J_3$) for $J_1$=+1.0, $J_2$=-1.0, Δ=0.0; (d) in the plane (Δ, $J_2$) for $J_1$=+1.0, $J_3$=-1.0, H=+0.0 ; (e) in the plane (Δ, $J_3$) for $J_1$=+1.0, $J_2$=-1.0, H=0.0 and (f) in the plane (H, Δ) for $J_1$=+1.0, $J_2$= $J_3$=-1.0.